\documentclass[reprint,amsmath,amssymb,aps,prstper,nofootinbib,superscriptaddress]{revtex4-2}
\usepackage{graphicx}
\usepackage{dcolumn}
\usepackage{bm}
\usepackage{hyperref}
\usepackage{xcolor}
\usepackage{comment}
\usepackage{rotating}
\usepackage{colortbl}
\usepackage{enumitem}

\usepackage{graphicx}
\usepackage{dcolumn}
\usepackage{bm}

\begin{document}

\title{Impact of Outreach on Physics Student Development: Quantitative Results from a National Survey}

\author{Jonathan D. Perry}
  \affiliation{Department of Physics, University of Texas at Austin, Austin, Texas 78712}

  \author{Toni Sauncy}
  \affiliation{Department of Physics, Texas Lutheran University, Seguin, Texas, 78155}
  \author{Susan White}
  \affiliation{American Institute of Physics, College Park, Maryland, 20740}
  \author{Rachel L Ivie }
  \affiliation{American Association of Physics Teachers, College Park, Maryland, 20740}
  \author{John Tyler}
  \affiliation{American Institute of Physics, College Park, Maryland, 20740}
  \author{Tatiana Erukhimova}
  \email{etanya@tamu.edu}
  \affiliation{Department of Physics \& Astronomy, Texas A\&M University, College Station, Texas, 77843}

\date{\today}

\begin{abstract}
This work reports on results from the first national study of the impact of student facilitation of informal physics outreach programs on their physics identity, sense of belonging, and essential career skill development. Drawing on results from studies at a single institution with a well-developed physics outreach program, a national survey was developed and distributed through the Society of Physics Students' network to more than five thousand individuals. Responses to closed-form questions on the survey were analyzed descriptively and using multiple regression analysis to evaluate the relationship between student participation in informal physics outreach programs and the constructs of interest. Results show a strong association between students engaging in outreach with their confidence in communicating their physics knowledge, the development of key career skills, and direct connections with growth mindset and sense of belonging. These results may be useful to physics and other STEM departments around the United States who are seeking to elevate and broaden the learning experience of students.

\end{abstract}

\maketitle


\section{Introduction}

Students entering an undergraduate physics program face many challenges to their success, among them both academic and social hurdles. In recent years, studies have expanded our awareness of how addressing social challenges, such as belonging and growth mindsets, can support students in persisting and anchoring their academics \cite{lewis2017fitting, yeager2016teaching, james2020time, burnette2023systematic, williams2021belonging}. Improving students sense of belonging, defined by the feeling of being valued, accepted, and feeling like a legitimate member of their chosen scientific discipline, can reduce attrition rates \cite{dobbins2020time, Lewis2016, walton2007question}. While classroom interventions in recent years have sought to support the social side of learning more directly \cite{dweck2017needs, yeager2022synergistic, hecht2022peer}, programs beyond the classroom can play a pivotal role in student development. Such activities include undergraduate research, teaching, extracurricular groups, and informal physics outreach programs. This work focuses on the latter of these opportunities. 

Informal physics outreach programs are defined broadly as a set of activities beyond the classroom which bring students, faculty, and staff at colleges, universities, and national labs into direct or indirect contact with members of the public to engage with physics. These programs can vary greatly in size and frequency, ranging from small star parties, to short videos shared on social media, teaching in prisons, to large festivals attended by thousands. Often these programs rely on the engagement of undergraduate students who help to create and deliver content to groups of all ages. 

Engaging in these programs can support the development of students' physics identity: interest and motivation, performance and competence beliefs, recognition, and sense of belonging \cite{hazari2020context}. Development of student disciplinary identity and motivational beliefs can potentially increase student resilience within a chosen STEM major \cite{Hazari2010}. Additionally, participating in co-curricular experiences that develop abilities beyond the focus of classroom and laboratory studies, including informal communication, teamwork, and creativity, have been identified by the American Physical Society (APS) and American Association of Physics Teachers (AAPT) as a high priority for physics programs to prepare students for 21st century careers \cite{Heron2016, smith2021informal}. These studies clearly indicate that impactful, engaging courses must be combined with experiences outside the classroom to better provide undergraduate physics students with the opportunity to see themselves as contributing members of the physics community.  

While informal physics outreach programs are common components of physics programs at a wide range of institutions, a broad investigation of the impact of such programs on the undergraduate student experiences that may be used to develop specific guidance about recommended practice has not been reported. This paper presents the results of the first nationwide study on the impact of informal physics programs on undergraduate students who plan/facilitate them. We were focused on experiences of undergraduate students from different institutions across the United States related to building their physics identity, retention and persistence, sense of belonging, and the development of 21st century career skills. 

\subsection{Background and Prior Work}
Our prior results from a pilot study done at Texas A\&M University showed that students who participate in the conception and execution (presentation) of informal physics outreach programs experience the growth of a sense of purpose, and that such experiences provide the critical context needed to motivate persistence in physics \cite{rethman2021impact, randolph2022female}. This is not surprising since through informal physics outreach programs undergraduate students are offered the opportunity to practice leadership, planning, design, working in teams, explaining physics concepts to others, which offers a creative environment \cite{rethman2021impact, Hinko2012}. It also provides a rich and highly contextual opportunity for students to develop physics identity, experience sense of belonging, and practice important career skills \cite{graur2018education, leshner2007outreach}. Initiated with research by Finkelstein and Hinko \cite{hinko2013impacting}, there is an increasing interest in the Physics Education Research (PER) community on understanding how student participation in these programs supports the development of a physics and STEM identity, enhances retention and persistence, and creates a feeling of community \cite{Fracchiolla2016, national2015identifying, noauthor_researchers_2017, national2009learning, rainie2015scientists, besley2018scientists, brownell2013science, prefontaine2018intense}. Growing research supports the idea that these outreach programs have a vital impact on the undergraduate students who help to conceive and carry out these programs \cite{rethman2021impact, leshner2007outreach, Hinko2012, randolph2022female, national2015identifying, perry2021comparing, finkelstein2008acting, Hinko2016, Fracchiolla2016, Fracchiolla2020}, and that this effect may be even more significant for groups underrepresented in physics. 

Prior research suggests that developing a physics identity can help students choose physics as a career and persist in the field. Efforts to develop this identity have the potential to increase retention rates among students, including those usually underrepresented in physics. Among the key factors influencing students' persistence in physics and other STEM fields may be strong discipline-based identity, combined with sense of belonging to the STEM community, integration within an academic department, getting an external recognition as a STEM professional, motivational beliefs to do physics, physics self-efficacy, and real-world experience \cite{Fracchiolla2016, hazari2013science, Sauncy2015, bunshaft2015focus, Quan2019, Perez2014, Sawtelle2012, Zwolak2017, thiry2011experiences, Kalender2019, Hyater-Adams2018, hyater2019deconstructing, Close2016, Irving2015}. 

Equally important is student exposure to teaching physics to populations outside the academic world – opportunities that are amply provided by informal physics programs \cite{Hinko2016, feldon2011graduate, drane2014students, otero2010physics, garrett2023broadening}. The flexibility of the program formats provides students with vast opportunities to explain physics concepts to the general public. This communication format could potentially foster greater ownership and create opportunities for transformational experiences \cite{rethman2021impact, randolph2022female, garrett2023broadening}. Recent research indicates that by facilitating informal physics programs, students can develop their communication and presentation skills, enhance their design and fabrication abilities, and gain teamwork experience, all of which prepare them for 21st-century careers \cite{Fracchiolla2016, rethman2021impact, randolph2022female, Heron2016, Hinko2016}. Self-reported data show a positive impact on students' physics identity, the development of their physics voice, networking within the department and the broader STEM community, and their sense of being valued and accepted as a result of facilitating physics outreach programs \cite{Fracchiolla2020, rethman2021impact, randolph2022female, perry2021comparing, garrett2023broadening}.

However, these perceptions have relied on data from a relatively small number of participants drawn from a limited range of outreach programs. There is a clear gap in the literature examining a broader population of undergraduate physics students from various institutions across the country. This work generates a national dataset enabling a leap forward in our understanding of the connection between student facilitation of informal physics programs with their development as physicists, their sense of belonging, and their acquisition of skills needed for 21st century careers which can help to transform undergraduate student experience. Our team developed a survey which was distributed to the national network of the Society of Physics Students (SPS). The goal of this survey was to measure students' perceptions of their physics identity, sense of belonging, mindset, career skill readiness, as well as sample information about students’ level of engagement with informal physics programs along with student demographics. The new survey instrument was created with open and close-ended questions. In this paper we present the construction, distribution, and analysis of the survey with focus on the closed questions. The open-ended questions analysis will be presented in the subsequent paper with the preliminary results published in the PERC proceedings \cite{perry2024exploring}.

\subsection{Pilot Study Results and Framework}
An initial phase of this work has been completed at Texas A\&M University \cite{rethman2021impact, randolph2022female, perry2021comparing, garrett2023broadening}. A mixed-methods study was performed on the impact of five informal physics programs run by the Department of Physics and Astronomy on a relatively large number of undergraduate and graduate students. The students were mostly from physics and engineering departments. These five outreach programs differed in scale and frequency, spanning from a large annual physics festival with thousands in attendance to smaller-scale events held year-round. Self-reported data on the dimensions of physics identity, sense of belonging, and development of 21st century skills were collected using a survey instrument (117 responses) and interviews (35) with current and former undergraduate and graduate students.

The results of the pilot study survey data showed that over 80\% of survey participants reported that participating in outreach had a positive impact on their ability to recognize connections between physics topics and on their overall depth of conceptual understanding in physics. More than 85\% of survey participants reported that participating in outreach had a positive impact on their teamwork skills and ability to network within the physics department. These skills foster the development of multiple 21st century career skills including communication, teamwork, and design. When asked about their confidence in choosing a physics major, almost half of the responding students reported increased confidence in their choice after participating in outreach \cite{rethman2021impact}. Women reported a statistically significant increase in confidence in their choice of a physics major after facilitating informal physics programs \cite{randolph2022female}.

The results from the pilot study survey informed the design of the new survey used in this study to collect a nationwide data sample. The survey questions were grounded in theoretical frameworks similar to those used in the creation of the pilot survey. These frameworks included situated learning theory to define learning \cite{lave1991situated}, transformative learning theory to explain how powerful learning occurs  \cite{Mezirow2009, kegan2018form} , as well as the Hazari et al. \cite{Hazari2010} model and the Dynamic Systems Model of Role Identity (DSMRI) to define identity \cite{kaplan2017complex}. In the next sections we’ll describe the survey development and distribution. 

\section{Methods}
 
\subsection{Survey Development}

To explore the relationships between student participation in informal physics outreach programs and constructs including physics identity, mindset, and self-efficacy, we created a new survey. The survey questions, available in the Supplemental Materials, were developed based on themes from our previous studies at a single institution \cite{rethman2021impact, randolph2022female, perry2021comparing, garrett2023broadening} as well as collaborative expertise from Texas Lutheran University, Texas A\&M University, the University of Texas at Austin and the Statistical Research group of the American Institute of Physics. Likert scale items specifically sought to collect students' perceptions of their motivation and interest, sense of recognition from self and others, performance and competence beliefs, sense of belonging, confidence, desire to persist, growth versus fixed mindset, and self-efficacy. Further items captured students' level of participation in informal physics events, sense of ability to use essential nondisciplinary, or career, skills (\textit{e.g.} communication, teamwork, networking), and demographic factors (age, gender identity, identification as a member of the LGBTQIA+ community, domestic or international student, ethnicity, neurodivergence, parents education, classification, and home institution). Students who responded as participating in outreach also received prompts for three open-ended questions. As this work focuses on quantitative results, discussion of open-ended questions will be omitted from this paper. 



\begin{table}
    \caption{\label{tab_demo} Demographics of student responses to the national survey. In some cases, percentages do not add to 100\% as respondents could select `I prefer not to respond'. For ethnicity, respondents were allowed to choose multiple answers.}
    \begin{ruledtabular}
        \begin{tabular}{ll}
            Demographic & Percent \\ \hline
            \textit{Gender} & \\
            Men & 41\% \\
            Women & 52\% \\
            Another Identity & 5\% \\ \\
            \textit{Classification} & \\
            Freshman & 15\% \\
            Sophomore & 17\% \\
            Junior & 29\% \\
            Senior & 39\% \\ \\
            \textit{Race or Ethnicity} & \\
            American Indian or Alaska Native & 1\% \\
            Asian or Asian American & 9\% \\
            Black or African American & 6\% \\
            Hispanic or Latino & 12\% \\
            Pacific Islander & 1\% \\
            White & 77\% \\ \\
            \textit{LGBTQIA+} & \\
            Yes & 27\% \\
            No & 64\% \\ \\
            \textit{International Student} & \\
            Yes & 11\% \\
            No & 89\% \\ \\
            \textit{Parents' Education Level} & \\
            Did not complete high school & 3\% \\
            High school diploma & 9\% \\
            Associates or Technical degree & 5\% \\
            Some college/university & 8\% \\
            Bachelor's degree & 31\% \\
            Graduate degree STEM & 13\% \\
            Graduate degree non-STEM & 13\% \\
            Professional degree & 18\% \\ \\
            \textit{Outreach Participation} & \\
            Yes & 71\% \\
            No & 29\% \\
            
        \end{tabular}
    \end{ruledtabular}
\end{table}

\begin{table*}
    \caption{\label{cfa_factors} The final factors produced from a confirmatory factor analysis using a Varimax rotation and Kaiser normalization.}
    \begin{ruledtabular}
        \begin{tabular}{ll}
            Factor & Items \\ \hline
            Motivation & It is important for me to study physics/astronomy.\\
             & I feel motivated to complete my physics/astronomy tasks. \\
             & I am more interested in the field of physics/astronomy than I was when I graduated high school. \\
             & I think about transferring to another major.\\
             & I believe earning a bachelor’s degree in physics/astronomy is a realistic goal for me. \\
             Belonging & I feel I am part of my academic department/college community. \\
             & I feel a sense of community with peers in my major. \\
             & I have felt socially isolated in my major. \\
             Competence & I could do an excellent job on assignments in physics/astronomy. \\
             & I could do an excellent job on exams in physics/astronomy. \\
             & I can operate physics/astronomy lab materials and equipment. \\
             & I am able to explain physics/astronomy ideas to people outside my discipline. \\
             Confidence & I can get full-time employment in my chosen field. \\
             & I am able to explain physics/astronomy ideas to people in my discipline. \\
             Recognition & I see myself as a physicist/astronomer.  \\
             & I believe others see me as a physicist/astronomer. \\
             Self-Efficacy & I usually do well in mathematics. \\
             & Mathematics is harder for me than many of my classmates.\\
             & I am just not good at mathematics. \\
             & I learn things quickly in mathematics. \\
             Mindset & The harder you work at physics/astronomy, the better you can be. \\
             & You can only learn so much physics/astronomy, and there is not much that can be done to really change that. \\
             & I appreciate when people give me feedback on my physics/astronomy work, even if it’s not all positive. \\
        \end{tabular}
    \end{ruledtabular}
\end{table*}
\subsection{Pilot Survey \& Distribution}

To gather a national sample, we partnered with SPS for distribution. Initially, the survey was distributed to members of SPS groups at three institutions, corresponding to the home institutions of authors TE, TS, and JP. Students were emailed directly by American Institute of Physics (AIP) with weekly reminders sent over a three-week period. A total of 101 responses were collected for this pilot survey distribution. After reviewing responses, the survey was determined to be robust.  Analysis of responses showed consistency with prior results and high Cronbach $\alpha$ values for items targeting single constructs. A minor edit was made to one item, out of 61, at the suggestion of a member of the research team, but this did not change the meaning of the item. 

For the second phase, the survey was sent to more than 5,500 student members of SPS chapters across North America. Periodic reminders were sent over the next 4-6 weeks, encouraging students to complete the survey. At the end of the collection period, a total of 704 responses were received. Representing a roughly 13\% response rate, this is the largest data set collected focusing on student engagement with outreach to date. A breakdown of the demographics of survey respondents is shown in Table \ref{tab_demo}.



\subsection{Confirmatory Factor Analysis}

In developing the survey instrument, we hypothesized that we were measuring several constructs: Physics Identity (comprised of subconstructs of Interest and Motivation, Competence Beliefs, Recognition, and Sense of Belonging), Persistence, Confidence, Self-Efficacy, and Mindset. To validate our constructs, we performed a confirmatory factor analysis \cite{brown2012confirmatory}. Items framed in a negative tone were inverted so scales matched positively framed items. As different items had different scales (4-6 Likert responses), each item was normalized to a 0-1 scale. Factors were identified using principal component analysis using a Varimax rotation and a Kaiser normalization to produce orthogonal constructs \cite{brown2009choosing, weber2022mathematische}. 


We conducted confirmatory factor analysis with three and eight constructs. For the confirmatory factor analysis, we extracted the constructs using principal component analysis, and we used Varimax rotation with Kaiser Normalization to produce orthogonal constructs.

The Self-Efficacy construct performed as expected, with all four components loading onto the same factor. The Growth Mindset construct resulted in two factors with two components loading together and one not loading with any other components. The components of the Physics Identity construct are loaded into five factors. The Recognition subconstruct loaded as expected, and some components of two of the subconstructs (Belonging and Confidence) loaded together; however, the two remaining factors consist of subcomponents of the Interest, Belonging, Persistence, Competence, and Confidence subconstructs. 

Our \textit{a priori} expectation was that the Physics Identity components would load together, with the same being true for the Self-Efficacy and Growth Mindset components. While the Self-Efficacy components did load together, they loaded with two components from the Competence subcomponent of Physics Identity. Four Belonging components loaded together. Of course, factor analysis is as much an art as it is a science. After significant discussion the consensus was to use seven factors. 


The final seven factors and their constituent items are shown in Table \ref{cfa_factors}. This set of factors omits a single construct from the original list: persistence. The persistence item is observed to load primarily with items related to Interest and Motivation. This is understandable as the desire to persist in undergraduate studies towards a degree may be closely related to a student's interest and motivation to learn more physics \cite{vazquez1997some, riegle2011wants}.

\subsection{Models}

To examine the relationship between participation in informal physics outreach programs and individual characteristics, we employed logistic regression analysis. Dimensions included in the models comprised three categories: respondent characteristics (RC), institutional characteristics (IC), and the factors (FA) defined in the previous section. RC included a binary on student age determined by being younger or older than 24 (corresponding to traditional versus non-traditional aged students), gender identity, classification, race or ethnicity (respondents selecting more than one response were grouped as ``two or more races''), orientation, and parents highest level of education. IC included whether their college or university was public or private, as well as the higher physics/astronomy degree awarded (Bachelor's, Masters, or PhD). Our regression models examining constructs relation to participation in outreach (PO) took the form:

\begin{equation}
    PO = f(RC, IC, FA).
\end{equation}

In addition to regressions where participation in outreach was the outcome variable, we ran additional models to examine relationships between physics identity, self-efficacy, and mindset. For each mode, we ran two regressions of the form:

\begin{equation}
    Factor = f(RC, IC, FA, [PO])
\end{equation}

where the square brackets around participation in outreach, [PO], indicate that regressions were run both with and without this independent variable.

In total, 15 regression analyses were performed where each had about 27 variables, totaling over 400 hypothesis tests. Unless otherwise stated, statistically significant results in the next section are presented only for instances where $p<0.01$. We set the rigorous alpha level to reduce the likelihood of Type 1 errors.

\section{Results}
In this section, we highlight results from the closed questions from the national survey on student impacts of facilitating outreach programs. As noted in Table \ref{tab_demo}, 71\% of respondents participated in either some (45\%) or most (26\%) physics or astronomy outreach events offered by their department. Broadly, students who participated or did not participate in outreach programs were equally comfortable approaching faculty in their departments for help when they didn't understand something.  When comparing their awareness of these events, students who had participated in outreach were slightly more likely to report how often such events were offered through their department, Table \ref{tab_deptsup}. This difference is statistically significant (Mann-Whitney U, p$<$0.01) but is fairly small. Students who participated in outreach reported favorably about their perceptions of support, with 35\% strongly agreeing and 41\% agreeing that these activities were supported by their departments.  

\begin{table}
    \caption{\label{tab_deptsup} Student responses to the prompt ``How often are outreach activities available through the physics/astronomy department?''. Responses are separated by those who reported participating, or not participating, in outreach activities.}
    \begin{ruledtabular}
        \begin{tabular}{lcccc}
             & Never & Rarely & Sometimes & Most of the time \\ \hline
             Participated & 1\% & 16\% & 47\% & 35\% \\
             Not Participated & 9\% & 20\% & 46\% & 26\% \\
        \end{tabular}
    \end{ruledtabular}
\end{table}

When asked about their comfort explaining topics from physics or astronomy to others both in and outside of the discipline of physics/astronomy, students who participated in outreach reported higher average scores, Table \ref{tab_explain}. Students who had engaged in outreach programs were more likely to report being able to explain physics/astronomy ideas to others both in (p$<$0.01) and outside (p$<$0.001) of their home departments. 

\begin{table}
    \caption{\label{tab_explain} Student responses to prompts ``I am able to explain physics/astronomy ideas to people in/outside of my discipline.'' Responses are separated by those who reported participating, or not participating, in outreach. Response choices were Strongly Disagree (SD), Disagree (D), Neither Agree Nor Disagree (N), Agree (A), and Strongly Agree (SA).}
    \begin{ruledtabular}
        \begin{tabular}{lccccc}
             & SD & D & N & A & SA \\ \hline
             \textit{In discipline} & & & & & \\
             Participated & 0\% & 2\% & 9\% & 57\% & 32\% \\
             Not Participated & 0\% & 5\% & 14\% & 58\% & 23\% \\ \\
             \textit{Outside discipline} & & & & & \\
             Participated & 0\% & 3\% & 8\% & 55\% & 33\% \\
             Not Participated & 0\% & 7\% & 18\% & 53\% & 22\% \\
        \end{tabular}
    \end{ruledtabular}
\end{table}

Student motivations for engaging in departmental outreach programs are shown in Table \ref{tab_motives}. Respondents were able to select multiple potential options, though the provided list, developed from the pilot studies, seemed to capture a significant majority of motivations for students. The most common motivations were finding the discipline fascinating, giving back to the community, and (unsurprisingly) fun. The next most common motivations were to get to know new people, learn new things, and enhance their disciplinary knowledge. This is consistent with responses in a separate question asking to what extent their physics/astronomy knowledge had improved due to their outreach activities, where 30\% reported ``a great deal'', while 66\% reported ``somewhat''. The least common motivations were seen to be students figuring out if the discipline is right for them and obtaining extra credit for a class. 

\begin{table}
    \caption{\label{tab_motives} Student responses to ``Why do you participate in physics/astronomy outreach activities? Please select all that apply.''}
    \begin{ruledtabular}
        \begin{tabular}{lc}
             Choice & Selected \\ \hline
             Because I want to learn new things & 57\% \\
             I think that physics/astronomy is fascinating & 76\% \\
             To help me determine whether & \\
             \quad physics/astronomy is the right major for me & 13\% \\
             To get to know people & 50\% \\
             Extra credit for a class & 10\% \\
             Outreach helps me understand & \\
             \quad physics/astronomy better & 55\% \\
             Because it is fun & 79\% \\
             I want to give back to my community & 70\% \\
             Other & 6\%\\
        \end{tabular}
    \end{ruledtabular}
\end{table}

Student perceptions of the frequency of their use of, and preparation for employing, various career skills relevant to those who graduate with physics degrees is shown in Table \ref{tab_skills}. These career skills represent a significant portion of skills identified as important to 21$^{st}$ century careers \cite{Heron2016} and reported as being commonly used by recent bachelor's degree holders in both engineering and computer science positions \cite{aps_stats}. Students reported relatively high levels of use for most skills, particularly communication, including verbal explanations and non-verbal communication (e.g. posters, videos), and being able to be creative. It is remarkable then that such high percentages of students ($>$90\%) reported developing their conceptual understanding, creativity, and networking with other students some or most of the time while engaging in outreach. Respondents also reported high levels of networking and connecting with others in the department, mainly other undergraduate students and faculty, with others such as staff or researchers being less common. Leadership opportunities were slightly less commonly reported, though students still rated being well prepared to employ these skills during outreach activities. The data indicate that students overwhelmingly feel somewhat or well prepared to employ all of the skills from Table \ref{tab_skills}  in their future careers after earning their degrees.

\begin{table*}
    \caption{\label{tab_skills} Student responses to prompts about how often they use, or how well prepared they feel to use, various career skills in physics/astronomy outreach activities. Blank entries are due to a corresponding question not being posed in the survey.}
    \begin{ruledtabular}
        \begin{tabular}{lccccccc}
             & \multicolumn{4}{c}{While Participating} & \multicolumn{3}{c}{Prepared} \\
            Skill & Never & Rarely & Sometimes & Most of the time & Not at all & Somewhat & Well \\ \hline
            Verbal Communication & 3\% & 14\% & 36\% & 46\% & 2\% & 36\% & 61\% \\
            Non-Verbal Communication & 7\% & 19\% & 36\% & 38\% & 5\% & 40\% & 55\% \\
            Storytelling & 7\% & 17\% & 41\% & 35\% & 7\% & 44\% & 49\% \\
            Team Leadership & 9\% & 22\% & 35\% & 34\% & 5\% & 36\% & 59\% \\
            Event Leadership & 11\% & 22\% & 32\% & 34\% & 6\% & 42\% & 52\% \\
            Design \& Building & 9\% & 24\% & 34\% & 32\% & - & - & - \\
            Creativity & 2\% & 8\% & 38\% & 52\% & 2\% & 35\% & 63\% \\
            Conceptual Understanding & 1\% & 10\% & 41\% & 49\% & - & - & - \\
            Networking (other students) & 1\% & 5\% & 37\% & 59\% & 1\% & 26\% & 73\% \\
            Networking (faculty) & 3\% & 14\% & 34\% & 49\% & 3\% & 33\% & 64\% \\
            Networking (other) & 6\% & 17\% & 30\% & 26\% & 9\% & 37\% & 54\% \\

        \end{tabular}
    \end{ruledtabular}
\end{table*}

\begin{figure*}
    \centering
    \includegraphics[width=0.95\linewidth]{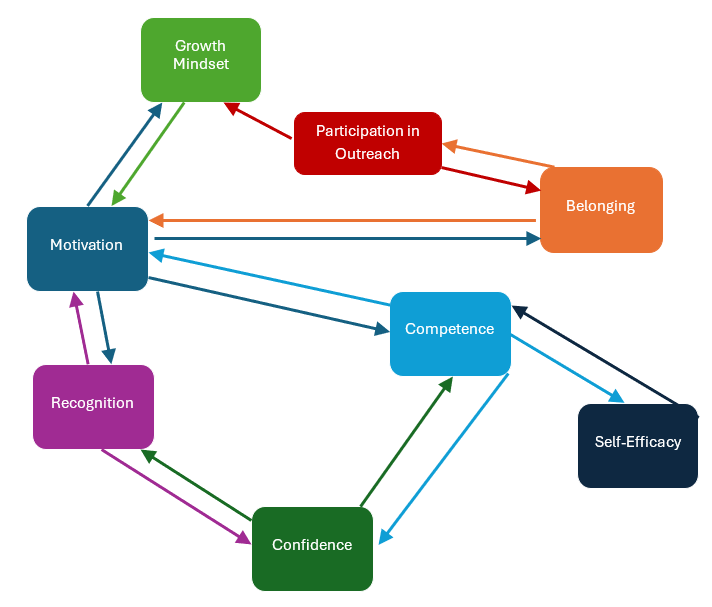}
    \caption{Visual representation of statistically significant relationships between factors and outreach participation. Arrows point towards the outcome variable for a given regression model for a significance level of p$<$0.01. }
    \label{regression_results}
\end{figure*}

The broad relationships between participation in outreach and the identity factors produced by factor analysis are shown in Fig. \ref{regression_results}. Below, we detail results for regression models first focusing on models with participation in outreach as the outcome variable, followed by all other models. The baseline comparison for hypothesis testing was a domestic, traditional aged, white, male senior, at a PhD granting institution, with at least one parent with a bachelor's degree. This baseline was selected as it represented the most common demographics of respondents to the survey.

\subsection{Participation in Outreach}

Regression models showed student participation in outreach is statistically significantly connected directly to both growth mindset and a sense of belonging. These relationships were observed to be positive, with respondents showing greater sense of belonging or a more growth mindset towards their physics intelligence and abilities when having participated in outreach. Though regression models do not provide information about causal direction, that is whether a greater sense of belonging led to participating in outreach or the reverse, the relationships between participation in outreach with belonging and mindset were strongly positive. 



Unsurprisingly, seniors were the most likely to have participated in outreach. From the regression models, first year students were five times less likely ($p<0.01$) to have participated in outreach. Sophomores ($0.05<p<0.10$) and juniors ($0.01<p<0.05$) were also observed to have been less likely to have participated in outreach compared to seniors, though not at a significant level for this study. 

In addition to positive and negative correlations, it is important to note which factors did not show a statistically significant difference in participation in outreach. Students who did not identify as men were no more or less likely than men to have participated in outreach. The same lack of difference is also observed for non-white respondents compared to white respondents, international compared to domestic students, first generation compared to continuing generation, and LQBTQIA+ identifying individuals compared to the rest of the sample. 

\subsection{Factors}

Regression models using factors from Table \ref{cfa_factors} were run both with and without the participation in outreach factor. Each factor was observed to be strongly linked with one to four other factors, whether outreach was included in the model or not. Only statistically significant relationships are described in this subsection. All other tests not mentioned upheld the null hypothesis of no difference. 

Motivation was positively related to belonging, competence, recognition, and mindset. No demographic factors were positively or negatively related to motivation. These results are identical whether participation in outreach was included or not. 

Belonging was positively related to motivation and participation in outreach, as noted above. Students enrolled in a department where the highest degree offered is a bachelor's degree were also more likely to report a higher sense of belonging compared to those in a PhD granting department. 

Competence was positively related to self-efficacy, motivation, and confidence. These results held whether participation in outreach was included or not. Women's responses indicated that they felt less competent compared to men. 

Recognition was positively related to motivation and confidence. Confidence was also positively related to recognition and competence. These results were independent of whether participation in outreach was included in the model. 

Self-efficacy was observed to be positively related only to the construct of competence, whether participation in outreach was present or not. Additionally, it was observed that juniors' self-efficacy scores were, on average, lower than the scores for seniors. No statistical difference was noted for other classifications. 

Mindset was observed to be positively related to motivation and participation in outreach. Additionally, first year students were more likely to exhibit a greater growth mindset. This was true whether participation in outreach was included or not. This is an unexpected result. As noted in the prior section first year students are less likely to participate in outreach, and those participating in outreach are more likely to exhibit a greater growth mindset. Combined these three pieces of information initially appear somewhat contradictory.


\section{Discussion}
In expanding beyond single institution studies, we have created, distributed, and analyzed a survey with a national sample of undergraduate physics students to explore the impact of informal physics outreach programs on their development. The data that we obtained could be helpful for physics departments around the country to improve the undergraduate student experience and programmatic outcomes. As seen in Table \ref{tab_explain} students who engaged in outreach were more comfortable discussing disciplinary ideas with those both in and out of their field. This signifies an enhanced confidence in their knowledge and their communication of that knowledge to both physics and non-physics audiences. This may be due, in part, to the fact that informal physics outreach programs are intrinsically active learning environments for students beyond the formal structure of a course allowing them to refine their ideas and take ownership of their knowledge. 

Results from Table \ref{tab_skills} show that informal physics outreach programs allow for the development of a broad set of essential career skills, which are not typically incorporated into traditional physics courses. Outreach programs vary in size, frequency, and focus but often rely on undergraduate student efforts to plan, set-up, populate with demonstrations or activities, and of course deliver to highly varied audiences. This provides a social and unstructured landscape allowing students to take on a variety of roles and duties which lend themselves to the development of these skills beyond the formal classroom. These skills were identified as essential for 21$^{st}$ century physics careers, in part by surveying employers who hire physics bachelor's \cite{Heron2016}. Results are also similar to our prior, single institution studies where students reported repeatedly about the impact of outreach on their communication, design, teamwork, leadership, and networking \cite{rethman2021impact, randolph2022female}. Given that physics courses are typically centered around development of understanding of physical principles, associated problem-solving skills, and lab procedures, it is no surprise to recall how rare opportunities are for students to use communication, design, and leadership skills. The data collected show that students engaging in informal physics programs receive significant opportunities to use these skills and feel very well prepared to employ them.



As this is a new survey, in discussing our results we will split our focus into two categories. First, we will contextualize results compared to prior literature of social constructs to note similarities and differences with existing knowledge. Second, we will interpret the unique knowledge from the regression models focused on the role of student participation in informal physics outreach programs. 

When we evaluated the results of this survey, we found that there is broad similarity with the results with prior research from physics and STEM education. Physics identity can play an important role in student development and persistence within the field. Results from our regression models show that the constructs related to physics identity, including motivation, recognition, competence, and sense of belonging are all interrelated. For example, students reporting a higher competence score are also more likely to report a higher motivation. This means that reporting a higher measure in one construct in the model could predict a higher outcome in other constructs. The interrelatedness of these factors is similar to what has been observed in other studies of physics identity, supporting the validity of the measures from our sample \cite{cheng2018examining, hazari2020context, rethman2021impact, randolph2022female}. 

Unsurprisingly, growth mindset and belonging are also both connected to student motivation. Studies from math and computer science have noted the strong link between a sense of belonging and student interest in continuing their major \cite{good2012women, master2016computing}. This link has also been noted for mindset in prior studies where interventions boosted measures of student interest and motivation in secondary science students in the UK and undergraduate computer science students in the US \cite{bedford2017growth, burnette2020growth}. It may be that course work alone is not be enough to impart such a growth mindset, which has been observed to stagnate or decrease across introductory courses \cite{marshman2018longitudinal}. Students with a growth mindset about their discipline are more likely to persist and overcome challenges, inevitable in their physics or science careers \cite{kalender2022framework, kim2022effects}.   


Curiously, measures of self-efficacy and competence beliefs did not depend on gender or ethnicity. It has often been observed in prior research that women and students from underrepresented groups in physics exhibit lower measures of these constructs across multiple studies \cite{marshman2018female, nissen2016gender, Sawtelle2012, kalender2017motivational}. One potential explanation for this difference between results could be due to the focus of the survey itself, which was about students' general feelings towards physics without reference to a classroom context. Indeed, this phenomena has been observed broadly for STEM students in a study by Wilson et al. \cite{wilson2015differences}. They noted that STEM self-efficacy was higher when measured generally about a discipline than for specific courses within a STEM discipline. This was true particularly for women and students from underrepresented groups \cite{wilson2015differences}. 


Most importantly, student participation in outreach was positively associated with students reporting more of a growth mindset and/or a greater sense of belonging within the discipline. This result with mindset is similar to our work in Randolph et al. \cite{randolph2022female} which showed that growth mindset was an important factor for the experience of women in physics at a single institution with a well-developed outreach program. Similar to mindset, students' sense of belonging can also have a significant impact on persistence within the field of physics \cite{Lewis2016}. Tests of interventions seeking to promote belonging have noted significant improvement, particularly for students from traditionally underrepresented groups in physics \cite{yeager2016teaching}. Those who feel like outsiders have been seen to leave physics at higher rates \cite{barthelemy2022lgbt+}. This again agrees with a result from our prior work in Rethman et al. \cite{rethman2021impact} connecting student facilitation of outreach with a higher sense of belonging to the STEM community. 

As stated earlier, there are two possible interpretations of the result linking participation in outreach with growth mindset and belonging. The first is that student engagement in informal physics outreach programs contributes to their sense of belonging within their physics community and their resilience to failure and their belief in the ability of their physics intelligence to improve over time. The second, of course, is that students with a higher growth mindset and sense of belonging are more willing and comfortable putting themselves forward to bring physics to the public and are those more likely to volunteer for these opportunities. We cautiously favor the first explanation, as Fig. \ref{regression_results}, shows one asymmetry between mindset and participation in outreach at the p$<$0.01 level. The reverse arrow, omitted in the figure, is significant at the p$<$0.05 level. This presents an interesting asymmetry showing that participation in outreach is more predictive of a higher growth mindset. We plan to explore this more in a subsequent study.

Informal physics outreach programs may support the development of growth mindset and belonging simply through their nature of being designed to provide opportunities for students to demonstrate their knowledge in low stress and fun environments. Far from the stress of homework and grades, students are able to show off their passion and knowledge of physics, often in ways that may reinforce that learning \cite{garrett2023broadening}. Also, outreach programs provide authentic opportunities for students to connect with fellow students, faculty, and researchers; as also noted in Rethman et al. \cite{rethman2021impact}. These programs also provide a rich playground for students to take ownership of projects, being the lead on demonstrating or even creating experimental apparatus that can be shown to dozens, or up to thousands, or members of the public. Such programs often bring together teams of students as well, fomenting their camaraderie and teamwork with fellow physics students \cite{rethman2021impact}.

It is interesting to note that for all the demographics analyzed in the regression models, only being a first year student was seen to have a significant, negative, impact on students likelihood of participating in outreach. While discussion of null results is uncommon, Conlin et al. make the case that discussion of such results can be beneficial to broadening our understanding of physics education \cite{conlin2019null}. Despite the persisting gender imbalance and underrepresentation of multiple ethnicities in physics \cite{aps_stats}, students who do not identify as men are no more or less likely to engage in outreach than men, and there are no statistically significant disparities in participation due to race or ethnicity. Though these data are insufficient to proclaim outreach as a great equalizer in engagement across demographic factors in physics, it is a positive signal that such programs appear to invite and benefit from the participation of all physics students.

\section{Limitations}

While results from the national survey demonstrate interesting relationships, we must note some limitations in our results. All data collected were self-reported from students responding to the survey. Further, due to the nature of hypothesis testing, there is potential that a small portion of the statistically significant (or not) results may be due to Type 1 (rejecting a true null hypothesis) or Type 2 (failing to reject a null hypothesis) errors. This may also be influenced by sparsely populated demographic categories in the sample. The purely quantitative nature of the results presented here lack nuance as to the experiences of students engaging in informal physics outreach programs. This will be the focus of a subsequent paper analyzing responses to the open-ended questions included in the survey.   

\section{Conclusion}
This work reports on the development, distribution, and analysis of the first national survey of the impacts of student participation in informal physics outreach programs. Drawing on our expertise, we developed a survey to measure constructs of physics identity, sense of belonging, and mindset which were distributed to the national SPS network of more than 5,500 students. Students who engaged in outreach programs reported being more comfortable explaining ideas from physics/astronomy to others both in and out of the discipline. A majority of participating students reported using essential career skills such as communication, networking, design, leadership during informal physics outreach programs and feeling prepared to use them. Students who participated were also slightly more likely to see outreach events being offered regularly from their departments, potentially due to their engagement generating more awareness. 

Results from the survey broadly mirror prior studies on physics identity and offer new knowledge about the potential benefits of outreach programs on university student development. These results provide important insights on how constructs which can be highly impactful to the student experience are related to student participation in informal physics outreach programs which are run by many departments throughout the United States. In particular, participation in outreach was observed to be positively correlated with students reporting a higher growth mindset and a greater sense of belonging. These results can potentially serve as a helpful justification for departments looking to develop, expand, or implement informal physics outreach programs, which do not necessarily have high capital costs to improve their student outcomes without changing their curricula. Such changes may be highly equitable across all student demographics, as only being a first year student was negatively correlated with likelihood to participate in outreach programs. No differences were observed based on gender or ethnicity. 

Within this work we have presented quantitative results from the national survey, without reference to responses to open ended questions. Analysis of the open ended questions will be the focus of a subsequent study, which may contain sufficient nuance and detail to suggest some directionality on the statistically significant relationships noted in this work. 

\section{Acknowledgments}

This work was supported by NSF Grant No. IUSE 2214493. We would like to thank Brad Conrad for his assistance in distributing the survey.


\bibliography{references}

\end{document}